# Medical Image Generation using Generative Adversarial Networks

Nripendra Kumar Singh and Khalid Raza*
Department of Computer Science, Jamia Millia Islamia, New Delhi-110025
*kraza@jmi.ac.in

May 19, 2020

**Abstract:** Generative adversarial networks (GANs) are unsupervised Deep Learning approach in the computer vision community which has gained significant attention from the last few years in identifying the internal structure of multimodal medical imaging data. The adversarial network simultaneously generates realistic medical images and corresponding annotations, which proven to be useful in many cases such as image augmentation, image registration, medical image generation, image reconstruction, and image-to-image translation. These properties bring the attention of the researcher in the field of medical image analysis and we are witness of rapid adaption in many novel and traditional applications. This chapter provides state-of-the-art progress in GANs-based clinical application in medical image generation, and cross-modality synthesis. The various framework of GANs which gained popularity in the interpretation of medical images, such as Deep Convolutional GAN (DCGAN), Laplacian GAN (LAPGAN), pix2pix, CycleGAN, and unsupervised image-to-image translation model (UNIT), continue to improve their performance by incorporating additional hybrid architecture, has been discussed. Further, some of the recent applications of these frameworks for image reconstruction, and synthesis, and future research directions in the area have been covered.

## 1. Introduction

Medical imaging plays a pivotal role in capturing high-quality images of almost all the visceral organs such as the brain, heart, lungs, kidneys, bones, soft tissues, etc. For image acquisition, there is a plethora of techniques used by a variety of imaging modalities including ultrasonography, computed tomography (CT), positron emission tomography (PET), and magnetic resonance imaging (MRI). However, the basic principles behind every modality are different for image acquisition, data processing, and complexity (Wani & Raza, 2018). For example, images from CT, PET, and MR images differ from each other in terms of complexity and dimensionality to incorporate modality-specific information which ultimately assists in better diagnosis. However, these diversities create a major constrain when it comes to cross-modality image synthesis. For instance, hybrid imaging involves simultaneous imaging from two modalities like MRI/PET, CT/PET imaging. The extraction of information of one modality from the hybrid images is usually a tough exercise.

For automated analysis of the medical image, it needs to meet certain criteria such as high-quality images, preserved low and high-level features, etc. A framework that performs translation of images from one modality to the other can be very promising by eliminating the need for multimodality scanning of the patient and help reduce time and expenditure. Generative adversarial network (GAN) is one such unsupervised framework that has carried out cross-modality image synthesis with significant accuracy and reliability (Raza &Singh, 2018).

This chapter is organized in the four sections and a discussion at the end. The first section is an introduction, the second section describes a brief background of GAN and its variant widely used the reconstruction and cross-modality translation. In the third section, we elaborate on some powerful variants of the GAN framework which got popularity in the medical image generation with desired output image resolution and cross-modality image-to-image translation. The fourth section contributes application of GAN in the medical image reconstruction and synthesis of different modality.

## 2. Generative Adversarial Network

The Generative Adversarial Networks (GAN), introduced by Ian J. Goodfellow and collaborators in 2014 (Goodfellow et al., 2014), is one of the recently developed approaches to 'generative modeling' using a flexible unsupervised deep learning architecture (Raza & Singh, 2018). Generative modeling is an important task carried out using unsupervised learning to automatically generate and learn the patterns in input data so that the model can be utilized for generating new examples (output) that could have been drawn using the original dataset. The ability of GAN to generate new content makes it more popular and useful in real-life image generation.

**Vanilla GAN**

Vanilla GAN is the initial variant for the synthesis of artificial images proposed by Goodfellow and colleagues (Goodfellow et al., 2014). The building block of this network, shown in Figure 1, is generative ($G$) and discriminative ($D$) models which uses fully connected layers of the neural network, due to this it has limited performance. It performs a generative approach for generating samples directly taking input $z$ from noise distribution $p(z)$, without any conditional information. The output of the generator is $x_g \sim G(z)$ and at the same time, instance of the real distribution $x_r \sim p_{data}(x)$ input to the discriminator model $D$, which produces single value output indicating the probability of generated sample is real or fake. In the case of the real image, the generator gets a reward in terms of positive gradient learning from discriminator and punishes the generator when the sample (image) is not close to real. However, the objective of $D$'s is like a binary classifier to distinguish pair of real or fake samples, while generator $G$ is trained much to generate a variety of realistic samples to confuse the discriminator model. The objective of competing models $G$ & $D$ shown in the following mathematical representation (Goodfellow et al., 2014),

$$\min_G \max_D V(D, G) = E_{x_r \sim p_{data}(x)}[logD(x_r)] + E_{x_g \sim p_z(z)}[1 - logD(G(z))] \quad (1)$$

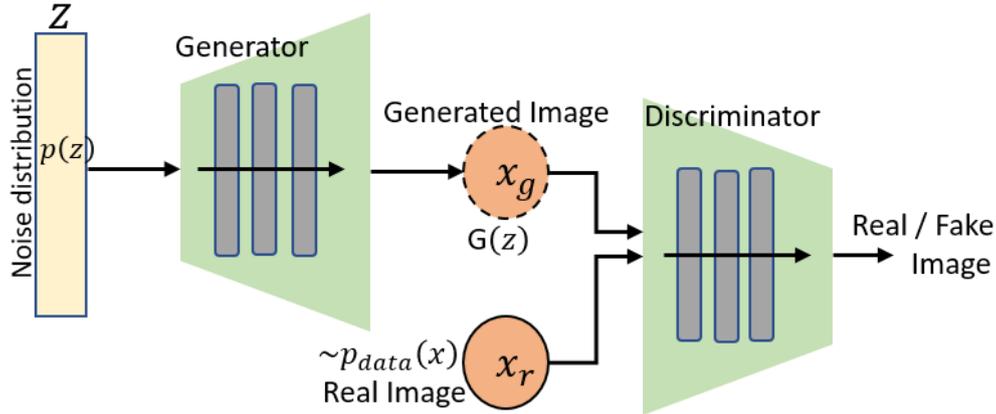

**Figure 1**. Typical structure of vanilla GAN

The challenges in GANs optimization are as follows:

*(i) Mode collapse*: A complete mode collapse is not arising commonly but happens partially, which is the most difficult and unreliable problem to solve in GAN. As the name suggests it is a situation when the generator continually produces a similar image and discriminator unable to consider the difference in generated samples, in this way generator easily fools the discriminator. It restricts the learning of the generator and focuses on a limited set of images instead of producing verities of images. The problem of collapse might be arising in inter-class or intra-class mode collapse.

*(ii) Vanishing gradients*: When discriminator becomes too strong, then in that situation discriminator does not provide enough corresponding gradients as feedback to the generator (when gradient loss function $(1 - logD(G(z)) \approx 0$, and learning of the generator stopped) and finally generator is unable to produce a competitive sample. Another condition is that when D's become too weak, then loss represents nothing to the generator.

*(iii) Problem with counting*: Early GANs framework was unable to distinguish several objects at particular location in generated image, and sometime GAN produce more or lesser number of nose holes, eye's and ear's even fails to place the right location.

*(iv) Problem with perspective*: Sometime GANs are unable to understand the perspective and failed to differentiate front view and back view objects, hence GANs could not work well at the conversion of 3D representation into 2D representation (Yadav et al., 2018).

## 3. GANs Framework for Medical Image Translation

In the initial setup, GAN was used as fully connected layers and no restrictions on data generation but, later on, it was replaced by fully convolutional down-sampling/up-sampling layers and conditional images constraints to get images with desired properties. Many different variants of the GAN framework was proposed to meet the desired output, which are Deep Convolutional Generative Adversarial Networks (DCGAN), Laplacian Generative Adversarial Networks (LAPGAN), pix2pix, CycleGAN, UNIT, CatGAN, BiGAN, InfoGAN, VAEGAN, StyleGAN and more. However, our main objective of this paper is to discuss GAN for image generation framework especially those which are popular among computer vision scientists for medical image-to-image translation. In this paper, we discuss DCGAN, LAPGAN, pix2pix, CycleGAN, and UNIT framework.

### 3.1 DCGAN

Deep Convolutional Generative Adversarial Networks (DCGAN) (Radford et al., 2016) produces better and stable training results when a fully connected layer is replaced by a fully convolutional layer. The architecture of the generator in DCGAN is illustrated in Figure 2(a).In the core of the framework, pooling layers replaced with fractional-stride convolutions that allowed it to learn from random input noise vector by own spatial upsampling to generate an image from it. There are two important changes adopted to modify the architecture of CNN are Batch Normalization (BatchNorm) and LeakyReLUactivation.BatchNorm (Ioffe & Szegedy, 2015)for regulating the poor initialization to prevent the deep generator from mode collapse which is a major drawback in the early GAN framework.LeakyReLU (Maas et al., 2013) activation introduced at the place of maxout activation (present in the vanilla GAN) all layers of a discriminator which improve higher resolution image output.

### 3.2 LAPGAN

It is very difficult to generate a high-resolution image directly from the noise vector. To solve this problem, Denton et al. (2015) proposed Laplacian Generative Adversarial Networks (LAPGAN) (Figure 2b) is a stack of conditional GAN model with Laplacian pyramid representation, each of layers adds higher frequency into a generated image. The merit of this model is successive sampling procedure used to get full resolution image, for the training, a set of generative models $\{G_0, G_1, \ldots G_k\}$ at each level k of the Laplacian Pyramid L(I) to captures the distribution coefficients $h_k$ for given input images $I$, $h_k$ is calculated by difference of adjacent levels in Gaussian Pyramid $G_k(I)$ and up-sampled value $U(I)$, i.e.

$$h_k = L_k(I) = G_k(I) - U(G_{k+1}(I)) = I_k - U(I_{k+1}) \qquad (2)$$

At the end of the level, it simply represents a low-frequency residual which is $h_k = I_k$.

Following equation is used for Sampling procedure with initializing $\tilde{I}_{k=1} = 0$,

$$\tilde{I}_k = U(\tilde{I}_{k=1}) + \tilde{h}_k = U(\tilde{I}_{k=1}) + G_k\left(Z_k, U(\tilde{I}_{k=1})\right) \qquad (3)$$

Figure 2(b) illustrates the sampling procedure for LAPGAN model. Suppose we need to sample a 64×64 image, for the number of layers K=3 required four generator models. Sampling started with a noise $z_3$ and generative model $G_3$ to generate $\tilde{I}_3$, which is up-sampled to the next level $l_2$ for the generative model $G_2$, . This process is repeated across successive levels to get the final full resolution image $I_o$.

### 3.3 pix2pix

The pix2pix is a supervised image-to-image translation model proposed by Isola et al. (2017) as depicted in Figure 2 (c). It has received a multi-domain user acceptance in the computer vision community for image synthesis, whose merit is to combine the loss of conditional GAN (CGAN) (Mirza & Osindero, 2014) with L1 regularizer loss, so that network not only learns the mapping from the input image to output image but also learn the loss function to generate the image as close to ground truth. The loss of CGAN is expressed as:

$$L_{CGAN}(G,D) = E_{x,y}[logD(x,y)] + E_{x,z}\left[log\left(1 - D(x,G(x,z))\right)\right] \quad (4)$$

Where $z \sim p(z)$ is random noise. The L1 regularization loss can be computed as,

$$L_{L1}(G) = E_{x,y \sim Pdata(x,y), z \sim P(z)}[||y - G(x,z)||_1] \quad (5)$$

The final objective function can be obtained by combining the above two equations as:

$$G^*, D^* = \arg\ \min_G \max_D L_{CGAN}(G,D) + \lambda L_{L1}(G) \quad (6)$$

Where $\lambda$ is a hyper-parameter coefficient introduced to balance the losses. Isola et al. (2017) propose two choices for training the pix2pix model, first for generator architecture based on U-Net(Ronneberger et al., 2015) have an encoder-decoder with skip connections (concatenate all channels between two layers), so it allows until a bottleneck layer to gather low-level information like the location of edges. The second scheme for discriminator architecture is PatchGAN (Li & Wand, 2016) which tries to classify N×N patches of images instead of the whole image.

### 3.4 CycleGAN

The cyclic adversarial generative network (CycleGAN) is proposed to perform higher-resolution image-to-image translation using unpaired data(Zhu et al., 2017). Figure 2(d) illustrates the architecture of CycleGAN, which preserves the history of input training image after the cycle of transformation and adding reconstruction loss. It consists of two generators, $G_{AB}$: transfer an image of the domain from A to B and another generator $G_{BA}$: doing the opposite transformation of $G_{AB}$. This cycle facilitates dual learning to the model. Also, the model consists of two discriminators, namely $D_A a$ and $D_B$ that decide the domain of an image. The adversarial loss function for $G_{AB}$ and $D_B$ pair is expressed as,

$$L_{GAN}(G_{AB}, D_B) = E_{b \sim P_B(b)}[logD_B(b)] + E_{a \sim P_A(a)}[1 - \log(D_B(G_{AB}(a)))] \quad (7)$$

And similarly, the adversarial loss for another pair $G_{BA}$ and $D_A$ is represented as $L_{GAN}(G_{BA}, D_A)$. Another loss is a cyclic-consistency loss to minimize the reconstruction error when an image in one domain translates to another domain and reverses back to the original domain. The cyclic-consistency loss is represented by the equation,

$$L_{cyc}(G_{AB}, G_{BA}) = E_{a \sim P_A(a)}[\|a - G_{BA}(G_{AB}(a))\|_1] + E_{b \sim P_B(b)}[\|b - G_{AB}(G_{BA}(b))\|_1] \quad (8)$$

After combining the above two equations (7) and (8) to obtain overall loss of the model, it would be,

$$L(G_{AB}, G_{BA}, D_A, D_B) = L_{GAN}(G_{AB}, D_B) + L_{GAN}(G_{BA}, D_A) + L_{cyc}(G_{AB}, G_{BA}) \quad (9)$$

The objective function of the CycleGAN can be represented by the following expression,

$$G_{AB}^*, G_{BA}^* = \arg\min_{G_{AB}, G_{BA}} \max_{D_A, D_B} L(G_{AB}, G_{BA}, D_A, D_B) \quad (10)$$

Even CycleGAN generates high-quality cross-domain medical images, pix2pix outperforms CycleGAN with a significant margin. Although, bothCycleGAN and pix2pix are unable to construct reverse geometric transformation (for example Zebra↔Horse). Visit CycleGAN project website (https://junyanz.github.io/CycleGAN/) for more examples and readings.

### 3.5 UNIT

An unsupervised image-to-image translation model (UNIT) was proposed by Liu and collaborators in 2017 (Liu et al., 2017). The model is hybrid in terms of the sharing of the weight of VAE (variational autoencoder) to coupled GAN (CGAN) ( Liu & Tuzel, 2016). Assumed that $x_1$ and $x_2$ be the same input image of different domains $X_A$ and $X_B$, then the encoders $E_1$ and $E_2$ share the same latent space, i.e. $E_1(X_A) = E_2(X_B)$. The UNIT framework is depicted in Figure 2(e). UNIT framework implements the shared-latent space assumption using weight sharing between last few layers of autoencoders and first few layers of generators. Due to shared-latent space, the objective function of UNIT is a combination of GAN and VAE objective function which implies the cycle-consistency constraints (Kim et al., 2017). Therefore, the result processing stream is called cycle-reconstruction stream, represented by the following mathematical equation,

$$\min_{E_1, E_2, G_1, G_2} \max_{D_1, D_2} L_{VAE_1}(E_1, G_1) + L_{GAN_1}(E_1, G_1, D_1) + L_{CC_1}(E_1, G_1, E_2, D_2)$$

$$L_{VAE_2}(E_2, G_2) + L_{GAN_2}(E_2, G_2, D_2) + L_{CC_2}(E_2, G_2, E_1, D_1) \quad (11)$$

Where $L_{VAE}$ represents objective for minimizing variational upper bond, $L_{GAN}$ is the objective function of GAN, and $L_{CC}$ is the objective function like VAE's to model the cycle-consistency constraint. Although, the UNIT framework performs superior to the CGAN ( Liu & Tuzel, 2016) on MNIST datasets, and there is no comparison available with another unsupervised model like CycleGAN (Zhu et al., 2017)

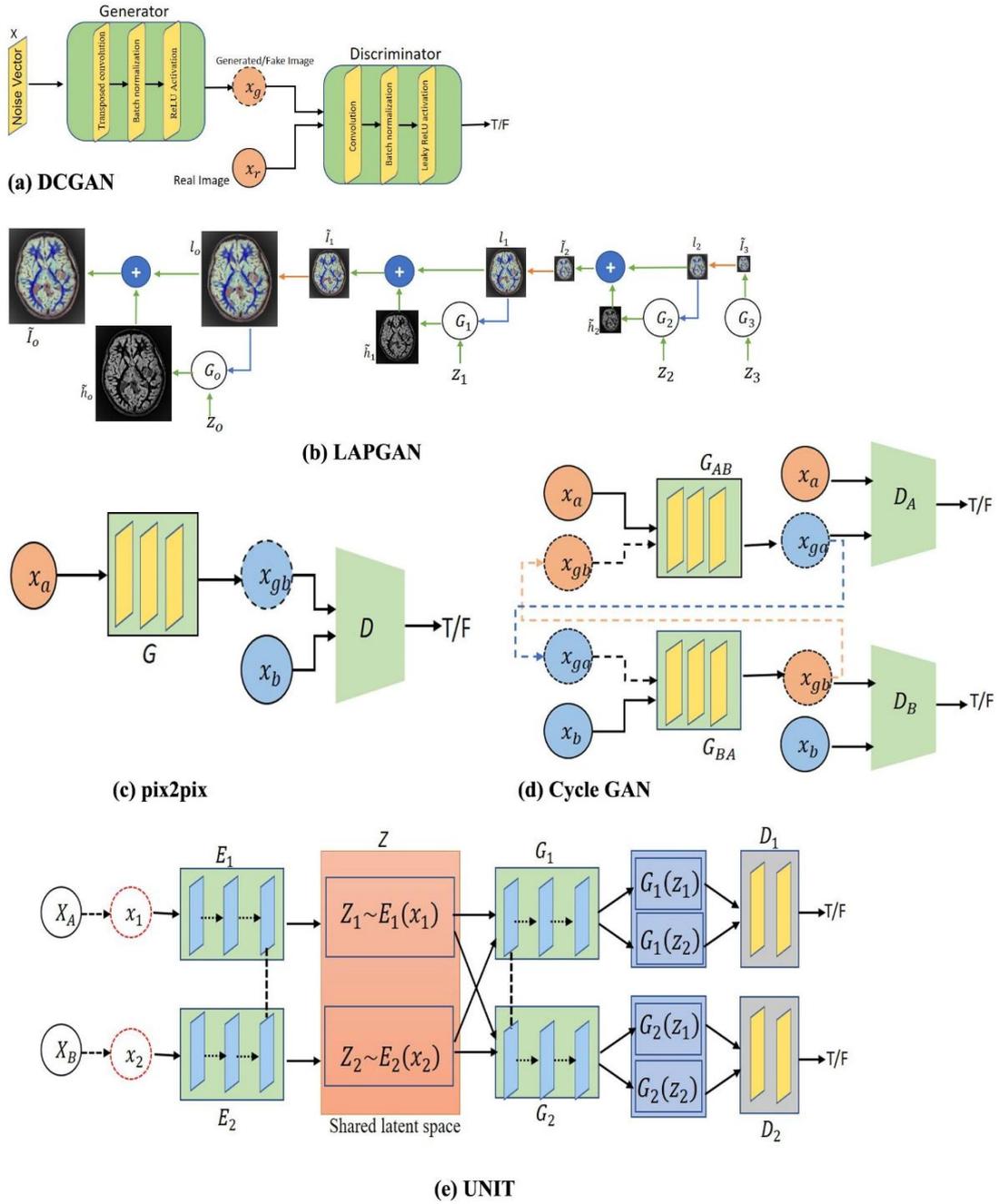

**Figure 2.** A schematic representation of a variant of GANs framework for image synthesis. (a) Represent architecture of DCGAN, where generator consist convolutional neural network followed by ReLU activation while discriminator uses another neural network followed by Leaky ReLU activation (b) Shows sampling procedure of LAPGAN (c)shows pix2pix framework takes input $x_a$ and $x_b$ as aligned training sample (d) shows the architecture of CycleGAN which takes input $x_a$ and $x_b$ as unaligned training sample (e) represents the architecture of the UNIT framework that contains two autoencoders with shared latent space.

## 4. Applications of GANs in Medical Imaging

Medical imaging exploits GAN in two different ways, which are generative and discriminative aspects. Generative aspects deal with the generation of new images using underlying structural information present in the training data. On the contrary, descriptive aspects of GAN learns structural feature from the natural or original images and rule out the anomalies present in the abnormal generated image. Most of the literature reviewed in this section has applied conditional GAN methods for image-to-image translation that suffers in certain forms such as undersampling, noise in the output image, and low spatial resolution. Figure 3 shows the example of GANs application for the generative and discriminative aspect of medical image generation.

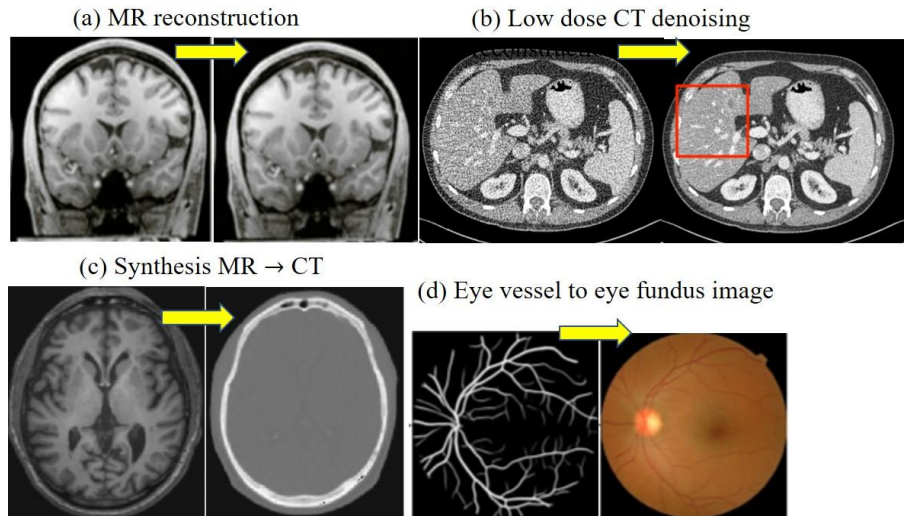

**Figure 3.** Application of GAN in medical image synthesis. All the figures are adapted from corresponding articles. (a) shows MR reconstruction from given reference image (Chen et al., 2018) (b) Low dose CT denoising (Shan et al., 2018) (c) shows input brain MRI used to generated equivalent CT image close to ground truth(Nie et al., 2017) (d) generation of synthetic eye fundus image from corresponding synthetic eye vessels(Costa et al., 2018).

**4.1 Reconstruction**

The medical image reconstruction is an essential step to obtain high-quality images for diagnosis with minimal patient discomfort, but the quality of such images limited by noise and artifact due to clinical constraints like the concentration of contrast media, amount of radiation administered to a patient in undergoing contrast MRI and PET respectively during the acquisition of images. Therefore, the objective to get reduced noise and another factor, analytical and iterative methods of reconstruction paradigm shifted to data-driven based machine learning methods (Armaniouset al., 2019b).

During literature survey we have observed that pix2pix and CycleGAN framework are frequently used in Magnetic resonance imaging (MRI) reconstruction (Shitrit & Riklin Raviv, 2017; Chen et al., 2018; Kim et al., 2018; Seitzer et al., 2018; Ran et al., 2019), low dose CT denoising (Wolterinket al., 2017; Shan et al., 2018; Yi & Babyn, 2018), optimization of the pre-trained network for sharpness detection and highlighting low contrast region in CT image (Wang et al., 2018), and synthesizing full dose equivalent PET from low dose concentration using skip connection introduced in the generator of conditional GAN (cGAN). Similarly, Liao et al. (2018) explore sparse view CBCT reconstruction for artifact reduction. They propose feature pyramid networks for the specialized discriminator and compute modulated focus map to reconstruct the output while preserving anatomical structure. Besides the reconstruction from lower sampling, it must ensure the domain data accuracy. MR reconstruction also imposes under-sample k-space data in the frequency domain (Mardani et al., 2019;

Quan et al., 2018; Yang et al., 2018).In image reconstruction, different types of losses have been used to distinguish local image structures, such as cycle consistency and identity loss together (Kang et al., 2019) in the denoising of cardiac CT. Wolterink et al. (2017) proposed low dose CT denoising after removing some domain loss but the final result compromised with local image structure. Reconstruction in MR is an exceptional case as it has a very much characterized forward and backward formula, for example, Fourier transformation. We summarised some studies related to medical image reconstruction in Table 1, and various losses to improve reconstruction accuracy described in Table 5.

**Table 1.** Summary of contributions in the medical image reconstruction over different modalities, the method for basic training network architecture, or with some modification with losses incurred (more description about losses described in Table 5) and the last column discuss remark for each contribution.

| Modalities | Methods | Losses | Remarks | References |
|---|---|---|---|---|
| MRI | Pix2pix | L1, 2 | Undersampled K-space reconstruction for accelerated MRI scan | Shitrit & Riklin Raviv (2017) |
| MRI | Pix2pix | L1, 2 | 3D super-resolution from single low-resolution input image using multi-level densely connected super-resolution network (mDCSRN). | Chen et al., (2018) |
| MRI | Pix2pix | L1,2 | Motion correction | Oksuz et al. (2018) |
| MRI | Pix2pix | L1,2,7,10 | Directly in complex-valued k-space data | Zhang et al. (2018) |
| MRI | Pix2pix | L1,2,11 | Undersampled K-space | Quan et al. (2018) |
| MRI | Pix2pix | L1,2,7,11 | Undersampled K-space | Yang et al. (2018) |
| MRI | Pix2pix | L1,2 | Super-resolution | Kim et al., (2018) |
| MRI | Pix2pix | L1,2,7 | Two-stage | Seitzer et al. (2018) |
| MRI | Pix2pix | L1,2,11 | Reconstruction into high-quality image Under sampled K-space | Mardani et al. (2019) |
| MRI | Pix2pix | L1,2,7,9 | Motion correction | Armaniouset al. (2019a) |
| MRI | Pix2pix | L1,2,7,9 | Inpainting | Armaniouset al. (2019b) |
| MRI | Pix2pix | L1,2 | Undersampled K-space | Ran et al. ( 2019) |
| PET | cGAN | L1,2 | 3D high-resolution image synthesizes equivalent to full dose PET image | Wang et al. (2018) |
| CT | Pix2pix | L1,2 | 3D denoising | Wolterink et al. (2017) |
| CT | Pix2pix | L1,2,7 | Sparse view CT reconstruction | Liao et al. (2018) |
| CT | Pix2pix | L1,2,5 | Denoising | Yi & Babyn, (2018) |
| CT | Pix2pix | L1,2,8 | 3D denoising | You et al. (2018) |
| CT | SGAN | L1,2,7 | Denoising, contrast enhance | Tang et al., (2018) |
| CT | Pix2pix | L1,7 | 3D denoising, transfer from 2D | Shan et al., (2018) |
| CT | CycleGAN | L1,2,10 | Super-resolution, denoising | You et al., 2018 |
| CT | CycleGAN | L1,3,12 | Denoising CT | Kang et al., (2019) |
| CT | Pix2pix | L1,2,7 | Denoising using the adjacent slice | Liu et al. (2020) |

**4.2 Medical Image Synthesis**

GANs framework provides a collective solution for augmenting training samples with sound results compared to traditional transformation, thus it is widely accepted for medical image synthesis and successfully overcomes the problem lacking in the volume of diagnostic imaging dataset of the positive or negative instance of each pathology. Another problem is lacking expertise in the annotation of diagnostic images which might be a big hurdle in the selection of supervised methods. Although, the multiple numbers of healthcare organization across the world and collaborative effort to build an open-access dataset of different modalities and pathology have been done, for example, The Cancer Imaging Archive (TCIA), National Biomedical Imaging Archive (NBIA), Radiologist Society of North America (RSNA) and Biobank. Researchers can access these image datasets with certain constraints.

*Unconditional Synthesis:* Unconditional synthesis simply generates an image from the latent space of a real sample before any conditional information. Commonly GAN models, DCGAN, PGGAN, and LAPGAN are adopted in medical image synthesis due to exceptional training stability. Where DCGAN generates limited output image quality, that could be up to 256×256 image size. However, DCGAN has been used to generating high-quality image samples of lungs nodule and liver lesion which easily deceive radiologists (Chuquicusma et al., 2018; Frid-Adar et al., 2018). Other methods used for generating higher resolution images are the iterative method, sharing weights among generators, but the hierarchical method may not do the same. Progressive Growing Generative Adversarial Network (PGGAN) performs the progressive growing techniques to get the desired realistic image. For example, Beers et al. (2018) can produce a synthetic image of MRI and retinal fundus of size up to 1024×1024 using the PGGAN model which is unprecedented for the previous model. The summary of articles related to unconditional synthesis is presented in Table 2.

**Table 2.** Summarize articles of unconditional synthesis of medical images

| Modalities | Methods | Remark | References |
|---|---|---|---|
| CT | PGGAN | Segmentation mapping using joint learning in augmenting brain image | Bowles et al (2018) |
| CT | DCGAN | Synthesizing liver lesion of each class using DCGAN, then classifying different class of lesion | Frid-Adar et al. (2018) |
| CT | DCGAN | Generate realistic lung nodule separately benign and malignant nodules | Chuquicusma et al. (2018) |
| MRI | Semi-Coupled GAN | Two-stage semi-supervised methods for detection of missing features from cardiac MR image | Zhang et al. (2017) |
| MRI | LAPGAN | Generating synthetic brain MR image slices | Calimeri et al. (2017) |
| MRI | DCGAN* | Semi-supervised achieve better than fully supervised learning with labeled and unlabelled 3D image segmentation | Mondal et al. (2018) |
| MRI | DCGAN | Manifold learning for image synthesis and denoising | Plassard et al. (2018) |
| MRI | PGGAN | Generating high-resolution Multimodal MR image of glioma and retinal fundus using progressive training, | Beers et al. (2018) |
| X-ray | DCGAN | Artificial chest X-ray augmentation and pathology classification | Salehinejad et al. (2018) |
| X-ray | DCGAN | Semi-supervised learning for abnormal cardiac classification | Madani et al. (2018) |
| Retinal | DCGAN | Vessel segmentation in the retinal fundus image | Lahiri et al. (2018) |
| Dermo | LAPGAN | Generating high-resolution skin lesion image | Baur & Navab (2018) |

*Conditional Synthesis:* Availability of right samples of medical imaging data is going to be a challenge especially when pathologies are involved. They rise two factors, like scarcity of the number of cases and large variation in anatomical location, appearance, and scale. Therefore, it is useful to synthesize artificially generated medical images in uncommon conditions by constraints on locations, segmentation maps or text, etc. Jin et al. (2018)augmented the lung CT data set with artificially synthesize nodules touching the lung border to improve pathological lung segmentation of CT. An adversarial autoencoder for a conditional segmentation map has been used to generate a retinal color image from a synthetic vessel tree which is a two-stage process (Costa et al., 2018). Moreover, generating brain MR by conditioned segmentation map used conditional GAN to learn automatic augmentation as well training samples for brain tumor segmentation (Mok & Chung, 2019), and trained CycleGAN network to correct geometric distortion in diffusion MR (Gu et al., 2019). Some contributions to conditional synthesis are summarised in Table 3.

**Table 3.** Summarize articles of conditional image synthesis of different modalities, representing modification in the given method either in network architecture or induced losses

| Modalities | Methods | Conditional Information | Authors |
|---|---|---|---|
| CT | Pix2pix | VOI with a removed central region | Jin et al. (2018) |
| MRI | CycleGAN | MR | Gu et al. (2019) |
| MRI | Cascade cGAN | Segmentation map | Lau et al. (2017) |
| MRI | Pix2pix | Segmentation map | Shin et al. (2018) |
| MRI | cGAN | Segmentation map | Mok & Chung (2019) |
| Ultrasound | Cascade cGAN | Segmentation map | Tom & Sheet (2018) |
| Retinal | cGAN | Vessel map | Zhao et al. (2018) |
| Retinal | VAE + cGAN | Vessel map | Costa et al. (2018) |
| Retinal | cGAN | Vessel map | Iqbal & Ali (2018) |
| X-ray | Pix2pix | Segmentation map | Mahapatra et al. (2018) |

*Cross Modality Synthesis:* Cross modality synthesis, for example, creating CT equivalent image dependents on MR images is esteemed to be helpful for different reasons. Consider a case study, when two or more imaging modalities say CT and MR require to provide supplementary information in diagnostic planning, in this case, the separate acquisition is required for the complete diagnosis, which increases the cost and time in the acquisition. So, cross-modality synthesis provides artificially generated samples of target modality from available modality with preserving anatomical structure or features, without separate medical image acquisition. Most of the methods are used in this section are similar to previous sections such as the CycleGAN-based method used where registration of images is more challenging. The pix2pix-based method is another well-accepted model used where imaging data registration ensures the data fidelity. Articles related to cross-modality synthesis summarized in Table 4. Summary of various losses used in the literature cited and introduced in different variants of pix2pix framework to get the desired output result is shown in Table 5.

**Table 4.** Summary of articles for cross-modality synthesis among different modality. The arrow→ represents one-way synthesis, and a dual arrow↔ represents two-way synthesis, * represent modification in the given method either in network architecture or induced losses.

| Modality | Methods | Losses | Remark | References |
|---|---|---|---|---|
| MR → CT | Cascade GAN | L1, 2, 4 | Context-aware network for the multi-subject synthesis | Nie et al. (2017) |
| MR ↔ CT | CycleGAN | L1, 3 | Unpaired training in synthesizing the cardiac image | Chartsias et al. (2017) |
| MR ↔ CT | Cycle GAN | L1, 3 | Training of unpaired 2D images to synthesis cross imaging modality | Wolterink et al. (2017) |
| MR → CT | cGAN | L1, 2 | Brain cancer analyzed to generate synCT | Emami et al. (2018) |

| MR ↔ CT | CycleGAN* | L1, 3, 6 | Generic synthesis of unpaired 3D cardiovascular data | Zhang et al. (2018) |
|---|---|---|---|---|
| MR ↔ CT | CycleGAN* | L1, 3, 4 | Unpaired training to synthesize musculoskeletal image | Hiasa et al. (2018) |
| MR ↔ CT | Pix2pix | L1, 2 | Paired training of 2D image to analyze prostate cancer for the complete pelvic region | Maspero et al. (2018) |
| MR ↔ CT | CycleGAN | L1, 3, 6 | Two-stage training and synthesis for abdominal image | Huo et al. (2019) |
| MR → CT | 3D U-Net | - | 3D patch-based network for Pelvic bone synCT generation | Florkow et al. (2020) |
| CT → MR | CycleGAN* | L1, 2, 3, 6, 7 | A two-step process to synthesis synMR in lung tumor segmentation | Jiang et al. (2018) |
| CT → MR | CycleGAN | L1, 2 | Both paired and unpaired training for brain tumor | Jin et al. (2019) |
| CT → PET | FCN + cGAN | L1, 2 | Synthesize paired training for liver lesion detection | Ben-Cohen et al. (2019) |
| PET → CT | cGAN | L1, 2, 7, 9 | Paired training for motion artifact and PET denoising | Armanious et al. (2020) |
| MR → PET | Cascade cGAN | L1, 2 | Brain anatomical feature from the sketch-refinement process used in the synthesis | Wei et al. (2018) |
| MR → PET | 3D Cycle GAN | L1, 2, 3 | Two stages paired training for Alzheimer's disease diagnosis | Pan et al. (2018) |
| PET → MR | Pix2pix | L1, 2 | Paired templet-based training for brain imaging data | Choi & Lee (2018) |

**Table 5.** Summary of various losses used in the literature cited and introduced in different variants of the pix2pix framework to get the desired output result.

| Abbreviation | Losses | Remark |
|---|---|---|
| L1 | $\mathcal{L}_{adversarial}$ | Discriminators introduced adversarial loss to maximize the probability of real or fake images. |
| L2 | $\mathcal{L}_{image}$ | Element wise loss in the structural similarity between real or fake in the image domain when aligned training is provided |
| L3 | $\mathcal{L}_{cycle}$ | Loss during cycle transformation to ensure self-similarity when unaligned training is provided |
| L4 | $\mathcal{L}_{gradient}$ | Loss in the gradient domain to focus on edges |
| L5 | $\mathcal{L}_{sharp}$ | Element-wise loss in which low contrast region computed to be image sharpness using a pre-trained network |
| L6 | $\mathcal{L}_{shape}$ | Shape loss is also to be segment loss in the reconstruction of the specified region |
| L7 | $\mathcal{L}_{perceptual}$ | Element wise loss to get a visual perception in a feature domain |
| L8 | $\mathcal{L}_{structure}$ | Structural loss is the patch-wise loss to get a human visual perception in the image domain |
| L9 | $\mathcal{L}_{style-content}$ | Style-content loss resembles style and content, where style is a correlation of low-level features |
| L10 | $\mathcal{L}_{self-reg}$ | Element wise loss in identifying among similar structure or in denoising in the image domain |
| L11 | $\mathcal{L}_{frequency}$ | Data fidelity loss in the frequency domain (K-space) especially in MRI reconstructions |
| L12 | $\mathcal{L}_{regulation}$ | Regulation loss in the generation of image contrast while keeping the balance across the row and column |

## 5. Conclusion and Future Research Directions

GANs framework has achieved great success in the field of medical image generation and image-to-image translation. We have discussed the weightiness of a significant rise in the study of medical imaging during the past 2-3 years. A detailed literature survey of GANs in medical imaging reported that about 46% of these articles are related to cross-modality image synthesis(Yi et al., 2019). A large section of research has focussed on the application of GANs in medical image synthesis of MRI imaging. The probable reason for the synthesis of MRI images is that it takes longer scan time for multiple sequence acquisition. Conversely, GAN effectively generates the next sequence from the acquired one, which saves time slots for another patient. The second reason may be the large number of MRI data set available in the public domain allowing researchers to have a surplus sample size for better model training. Further, a large fraction of studies conducted in the area of reconstruction and segmentation applications are due to better adversarial training and regulation on the generator's output of the GAN model for image-to-image translation framework. Although, conditional generation provides flexibility over augmentation and high resolution for training data. Some studies have synthesized chest X-rays for the classification of cardiac abnormalities and pathology (Madani et al., 2018; Salehinejad et al., 2018). While a very limited number of studies have been reported for the detection and registration of medical images.

Despite the diverse use of GANs, it has faced many challenges on the way for the adaptation of generated medical images directly into clinical diagnosis and decision making. Most of the work for image translation and reconstruction uses traditional methods of the metric system for quantitative evaluation of proposed methods. Especially when GAN incorporate additional loss, there arises difficulty in the optimization of the visual standard of an image in the absence of a dedicated reference metric. Recently, Armanious et al. (2020) have proposed MedGAN adopted perceptual study along with subjective measure, conducted by the domain expert (experienced radiologist) for an extensive evaluation of GAN generated image, but the downside is that it bears high-cost, time-consuming and difficult to generalize (Armanious et al., 2020). So, it is required to explore the validity of the metrics. Another big challenge is the absence of data fidelity loss in case of unpaired training. Therefore, it is unable to retain the information of the minor abnormality region during the cross-domain image-to-image translation process. Due to these problems, Cohen et al. (2018) suggested that the GANs generated image should not be straightaway used in clinical practice. Cohen et al. (2018) further experimented to confirm that unpaired training of CycleGAN subjected to bias in generated data from the target domain (Cohen et al., 2018). In a similar study of Mirsky et al. (2019) proven the possibility of intervening 3D medical imaging and bias only exists when the model was trained with normal standard data and tested with abnormal data (Mirsky et al., 2019).

Finally, we conclude that, even though there are many promising outcomes announced, the appropriation of GANs in clinical imaging is still in its early stages and further research in the area is needed to achieve a level of advancements essential for reliable application of GANs based imaging techniques in the clinical setup.

**Conflict of Interest**